# Electrolyte-gated organic synapse transistor interfaced with neurons

*Simon Desbief\*$, Michele di Lauro#, Stefano Casalini#$, David Guerin\*,*

*Silvia Tortorella+$, Marianna Barbalinardo+, Adrica Kyndiah+$, Mauro Murgia+,*

*Tobias Cramer+§, Fabio Biscarini# & Dominique Vuillaume\*£.*

\* Institute for Electronics Microelectronics and Nanotechnology (IEMN), CNRS, Avenue Poincaré, F-59652, Villeneuve d'Ascq, France.

# Life Science Dept., Università di Modena e Reggio Emilia, Via Campi 103, 41125 Modena, Italy.

+ Instituto per lo Studio dei Materiali Nanostrutturati (ISMN), CNR, Via P. Gobetti 101, 40129 Bologna, Italy.

§ University of Bologna, Dept. of Physics and Astronomy, Viale Berti Pichat 6/2, 40127 Bologna, Italy.

**ABSTRACT**

We demonstrate an electrolyte-gated hybrid nanoparticle/organic synapstor (synapse-transistor, termed EGOS) that exhibits short-term plasticity as biological synapses. The response of EGOS makes it suitable to be interfaced with neurons: short-term plasticity is observed at spike voltage as low as 50 mV (in a par with the amplitude of action potential in neurons) and with a typical response time in the range of tens milliseconds. Human neuroblastoma stem cells are adhered and differentiated into neurons on top of EGOS. We observe that the presence of the cells does not alter short-term plasticity of the device.

# 1. INTRODUCTION

There is a growing interest to emerging or alternative logic architectures beyond the Boolean architecture based on silicon CMOS technology (see a review in Ref. [1]). In this context, the organic synapse transistor (termed synapstor) has been initially designed and studied[2, 3] as the basic element for neuro-inspired computing architectures (namely artificial neural networks). These devices are based on the trapping/detrapping of charges by an ensemble of gold nanoparticles (NPs) placed at the gate dielectric/organic semiconductor (OSC) interface. The presence of NPs embedded in the OSC at the interface with the gate dielectric confers memory effects in a single transistor. For this reason, this device was also named Nanoparticle Organic Memory Field-Effect Transistor (NOMFET).[2, 3] Organic synapstor mimics the two main synaptic plasticity behaviors of biological spiking synapses[2, 3]: STP - Short Term Plasticity - and STDP - Spike Timing Dependent Plasticity. STP is characterized by the fact that the output signal of the synapse decreases with the number of spikes (depressing behavior) for spikes at the highest frequencies of operation, whereas a facilitating behavior (increase of the output signal with the number of pulses) is observed for the lowest operational frequency. The synaptic weight (i.e. the signal transmission efficiency) of the biological synapse is tuned by the density of input spikes (frequency coding of information). Depending on the frequency of a sequence of input spike voltages with respect to the typical charging/discharging dynamics of the NPs in the OSC channel, we demonstrated that the output drain current of the synapstor can be also modulated by the frequency of the input spikes.[2, 4] Thus, the transconductance of the organic synapstor mimics the STP behavior of the synapse. Recently, we have reported that organic synapstors can be optimized to work with low voltage spikes (1V), with a typical response time of about 100 ms and a low energy dissipation of about 2 nJ/spike.[5]

In the field of bioelectronics, organic transistors like electrolyte-gated organic field effect transistor (EGOFET)[6-12] and Organic Electro-Chemical Transistor (OECT)[13-16] have been interfaced with neuronal cells for biochemical signal recording and transduction of bio-electrical signals from cells and tissues.[17-24] Synaptic behavior (STP) has also recently been demonstrated for OECTs made with PEDOT:PSS films with a quite different principle since this STP behavior is driven by the dynamics of ions diffusion between the electrolyte and the PEDOT:PSS layer.[25, 26]

In this paper, we demonstrate electrolyte-gated organic synapstor (EGOS) with performances suitable to be interfaced to neurons. STP is demonstrated at spike voltage as low as 50 mV (in a par with the amplitude of action potential in neurons) and with a dynamic response down to tens milliseconds. We report the adhesion of Human Neuroblastoma cells (SH-SY5Y) and their differentiation into neurons on top of the EGOS. We demonstrate that EGOS STP is not altered throughout the cell growth and differentiation protocol. Interfacing neural cells to EGOS is attractive for synapse prosthesis application because the device could mediate the signal transmission between adjacent cells that are not connected through synapses. In this sense, the device might replace a chemical synapse with an electronic synapsis gated by a frequency dependent input supplied by the action potential of the transmitting neuron.

## 2. RESULTS AND DISCUSSION

### 2.1. EGOS without neurons

Figure 1 shows representative output characteristics $I_D$-$V_D$ of the devices measured in air (silicon bottom gate, NOMFET configuration) and in saline aqueous solution (Pt wire top gate, EGOS configuration). Devices have a channel length L=1-50 μm, channel width W=1000 μm, $SiO_2$ thickness 200 nm, Au NPs 10 nm in diameter, 40 nm thick pentacene, electrolyte 0.1 M NaCl in deionized water, see details in section "Methods". Noticeably, EGOS displays field-effect below 0.8V due to the high capacitive coupling between the OSC and the electrolyte, whereas the organic synapstor operating in air requires few tens V. We limited the voltage range below 0.8 V to avoid water electrolysis and faradaic current. These characteristics are similar to the ones already reported for electrolyte gated organic field effect transistor (EGOFET),[10, 11, 20, 27, 28] that consists of the same device structure as EGOS without the NPs embedded in the pentacene film. From the transfer $I_D$-$V_G$ curves in the saturation regime (not shown), we extracted the hole mobility for the EGOS assuming a gate-voltage independent double layer capacitance $C_{DL}$ of 14 μF/cm$^2$ as measured for pentacene EGOFET with a similar saline aqueous solution.[12] We obtain $\mu_{EGOS} \approx$ 1-5x10$^{-4}$ cm$^2$/V.s for devices with L between 1 and 50 μm (see Fig. S5 of SI). The hole mobility for the same device operated in air with the bottom silicon gate is 10$^{-2}$-10$^{-1}$ cm$^2$/Vs, as already reported for these organic synapstors.[5] This decrease of the hole mobility for EGOFET has already been observed.[20, 29, 30] A possible explanation was suggested[11] and related to the fact that the charge carrier channel is compressed at the surface of the organic semiconductor, i.e. at the interface with the electrolyte, due to the increased capacitance coupling in electrolyte-gated devices. Thus, the charge transport is more sensitive to surface defects and roughness. This strong decrease of the mobility is consistent with the fact that for the pentacene thickness used here (40 nm equivalent to a nominal coverage of about 25 MLs) the pentacene surface is quite rough due to the presence of NPs which induce pentacene films with a disordered morphology (the rms roughness is ca. 15 nm for the pentacene with NPs, it is ca. 8 nm for P5 alone, see AFM images in Refs. [2] and [5] also Fig. S6 in SI). This more disordered pentacene film in EGOS is also due to the layer to island growth transition of pentacene that occurs above a few monolayers.[5] For the EGOS on quartz (Fig. 1-c), we get an even lower mobility 0.7-2.0x10$^{-4}$ cm$^2$/Vs but with a larger dispersion (some of the devices exhibit charge mobility as high as 1-3x10$^{-3}$ cm$^2$/Vs).

Figure 2-a shows typical STP response of the EGOS excited by a series of spikes at different frequencies (50-500 Hz) and at $V_P$ = -50 mV. We clearly

observe the facilitating (increase of output $I_D$ spikes) and depressing (decrease of output $I_D$ spikes) behaviors when changing the frequency of the input spikes, as observed for a biological synapse.[31] A typical response of the corresponding NOMFET device in air (bottom silicon gate) is shown for comparison in Fig. 2-b in agreement with Ref. [5]. The main result is that almost the same STP behavior is observed for the EGOS at lower spike amplitudes (here $V_p$ = -50 mV) compared to the organic synapstor operated in air (here $V_p$=-1V). The STP response of organic synapstor is characterized by the charge/discharge time constant of the NPs embedded in the pentacene film.[2-4] This time constant is obtained by fitting (red squares in Fig. 2) an analytical and iterative model (see details in Refs. [2, 4]) on the drain current. The model accounts for the effect of charge (during the spike) and discharge (between two successive spikes) of the NPs (Coulomb effect) on the current. As shown in Fig. 2, STP time constants are in the same range (few tens ms) for the devices operated in air and in saline aqueous solution. This demonstrates that the dynamics arises from the NPs and not from the gate coupling ($SiO_2$ vs. electrolyte) which are strongly different ($C_{SiO2}$= 17 nF/cm$^2$ for the 200 nm thick $SiO_2$, $C_{DL}$ ≃ 14 µF/cm$^2$ for electrolyte). A slightly faster (a few ms, ca 5 ms) typical intrinsic response time has been measured for pentacene EGOFETs (without NPs) of the same size and same ionic strength of the electrolyte.[12] For EGOS on quartz, the device exhibits STP at lower frequencies (Fig. 2-c) with a STP time constant in the range of few seconds because the charge mobility is lower (see above) and the channel length longer (15 µm for EGOS on quartz, 1 μm for EGOS on Si/SiO$_2$, the STP time constant increasing with channel length as reported elsewhere[2]). In both cases (namely EGOSs on Si/SiO$_2$ and on quartz substrates), the density of NPs (see Figs. S1 and S3 in SI) is similar (1-5x10$^{10}$ NPs/cm$^2$).

The characterization of the synaptic-like STP behavior of the EGOS has been further extended by systematically varying the spike amplitude $V_p$. Figure 3 shows the variation of the STP amplitude $\Delta I/I$ vs $V_p$ (in abs. value) for EGOS (only on Si/SiO$_2$ substrates) and the synapstor operated in air ($\Delta I$ is the maximum drain current variation during a complete spike sequence, and I the average current for the complete sequence). For devices shown in Figs. 2-a and 2-b, $\Delta I/I$ is 0.32 (EGOS, $V_p$=-50mV) and 0.28 (organic synapstor in air, $V_p$=-1V). We observed that for the standard organic synapstor in air, the STP amplitude $\Delta I/I$ is increased with the amplitude of the input spikes. This is explained by the fact that the output drain current variation is dependent on the amount of charges stored in the NPs. In a model detailed in Refs [2, 3] the STP behavior of our devices is related to charge and discharge of carriers by the NPs embedded in the organic semiconductor layer and characterized by a typical charge/discharge time constant. When the period of the applied spikes is lower (higher, respectively) than this time constant, the device shows the depressing behavior (facilitating behavior, respectively). The output current is thus directly modulated by the amount of charges trapped in the NPs (Coulomb effect). This amount of trapped charges, $Q_{NP}$, in the NPs depends on the applied voltage pulse on the gate, $V_G$, through their capacitance C : $Q_{NP} = C_{NP} V_G$. Thus, the STP amplitude of the output current is expected to increase with the spike amplitude, as detailed in Ref. [3] (Eqs. S8 and S9 in SI of Ref. [3]). These equations resumes as $I_D=g_{D0} \exp(-\gamma C_{NP}V_G) V_D$, with $I_D$ the drain current (output), $V_D$ and $V_G$ are the drain and source voltages, $g_{D0}$ the output conductance with no charge in the NPs, $C_{NP}$ the capacitance of the NPs and $\gamma$ a constant (see Ref. [3]). This equation (for a single pulse on the gate) indicates that the drain current variations (related to the STP amplitude) are larger when increasing voltage (spike on the gate). Note that the experiments reported in Fig. 3 are more complicated since we applied a train of spikes and not a single pulse but the tendencies are captured as also shown by a more detailed modelization in Ref. [3]. Extrapolation below pulse voltage of 1V predicts the STP amplitude to be very weak (as shown Fig. 3) for synapstor in air. However, we demonstrate experimentally that EGOSs are still working with spike amplitude down to 50 mV with a STP amplitude $\Delta I/I$ up to 0.3.

**2.2. EGOS interfaced with neurons.**

According to the protocol described in section "Methods" and in the supporting information, Human neuroblastoma stem cells SH-SY5Y and NE-4C have been adhered, grown and differentiated into neurons on top of EGOSs. Every biological experiments were performed in parallel with reference samples, such as polystyrene petri dish and bare quartz substrates (data not shown). Firstly, in-vitro biocompatibility of pentacene/NPs has been assessed for both cell lines. In particular, these cells have been grown on bare quartz and test pattern (interdigitated Au electrodes), both coated with 15 nm of pentacene on NPs (see Fig. 4). On the bare quartz (Figs. 4a and 4b), SH-SY5Y shows a wider and denser network than NE-4C. The same result has been obtained on test patterns (see Figs. 4c and d), i.e. SH-SY5Y coverage on the interdigitated electrodes is greater than that of NE-4C. In order to maximize the surface coverage of the cells we decided to use only SH-SY5Y cells for the electrical measurements.

Figure 5 shows the optical images of the SH-SY5Y growth on EGOS at different days of the incubation. These images show the correct growth of a neuron network after 6-7 days. It is known that cells are extremely sensitive to their surroundings. As a result, their cytoskeletal organization, shape, motility and fate can be affected upon deposition on a solid substrate. We note that these results are similar as those already reported for pentacene EGOFET.[20, 32] This implies that the presence of the Au NPs embedded in the organic layer is not detrimental for the neuronal cell growth, albeit the hybrid NP/pentacene layer is rougher than the pentacene alone, with smaller pentacene grains as evidenced by AFM images (Fig. S6 in SI).[5] Although the interdigitated electrodes represent a relevant change of topography, still a substantial density of cells is present on the active area of the EGOSs. Immuno-fluorescence imaging of SH-SY5Y after staining of specific probes (Fig. 6) on the interdigitated electrodes shows the successful differentiation of SH-SY5Y cells into a neuronal phenotype (i.e. positivity to β-III Tubulin and MAP2). Fig. 6a shows the labelling of β-III tubulin, and Fig. 6b the labelling of MAP2 (microtubule-associated protein), specific of a differentiation in neuronal cells. The blue color (DAPI, Fig. 6b) indicates the nuclei of the cells. These results show that EGOS are biocompatible substrates towards these cells. Figures 4-6 illustrate quantitatively that the cells adhere to the surface, they grow and cover almost to confluence the electrodes in seven days, and they are differentiated into neurons. We have not carried out, in this case, a quantitative analysis, as this was reported in previous papers.[20, 32]

The electrical response of EGOS has been monitored before and after cell growth. Figure 7 shows the STP response of two devices recorded at different times during the cell growth. In these experiments, we have used a series of spike at the lowest frequencies (0.5Hz, 5Hz, 1Hz, 0.5Hz and 2Hz, spike width 100 ms) because of the lower mobility previously measured (see above and Fig. 2-c). The STP on device A has been recorded in the cell culture media at day 0 before cell seeding and day 1 after cell adhesion. The two responses slightly differ in terms of time constant and relative amplitude $\Delta I/I$ are: 4.7 s and 2.6 s and 0.43 and 0.35, for day 0 and day 1 respectively. The STP on device B has been recorded at day 0 before cell seeding and day 6 (i.e. with fully differentiated cells). The STP time constant remains in the same range (5.5 s and 4 s for day 0 and day 6, respectively), but the STP response amplitude is slightly lowered, namely from $\Delta I/I$ =0.44 to 0.30 after cell differentiation. The main feature is that the presence of neurons on the top of the EGOS does not affect drastically the STP behavior. We also note that the observed dynamics in these experiments is solely due to the charging/discharging of the NPs and not related to a possible electrical activity of the neuronal cells arising from the voltage spike stimulation. The electro-stimulated activity of neurons exhibits a much lower drain current level (few 10 nA) and fastest time scale (50-100 ms) as previously recorded with pentacene EGOFETs.[20]

3. CONCLUSION

In conclusion, we have shown that electrolyte-gated organic synapstor (EGOS) mimics the synapse plasticity (STP, short-term plasticity) at spike voltage as low as 50 mV and with a dynamic response in the range of few tens of ms (EGOS on Si/SiO$_2$ substrate). While this spike amplitude is on a par with the amplitude of the intracellular action potential, it is about 2 orders of magnitude smaller than the amplitude of the extracellular potential (few hundreds µV). We also reported the adhesion of stem cells and their differentiation into neurons on top of these hybrid nanoparticle/organic semiconductor devices fabricated on quartz substrate, and have demonstrated that STP of the EGOS is not strongly altered upon the cell growth and differentiation. This feature implies that the neurons are weakly coupled to EGOS. With the fact that the spike amplitude required to drive the EGOSs is lower than the extracellular action potential, these features may lead to the conclusion that it would be difficult to trigger the EGOS from the neuron signal. However, it was previously observed that synchronized action by a neuronal population is able to trigger a sizable change in EGOFETs made with the same architecture (without nanoparticles though).[20] Experiments with the cells triggering the EGOS are out of the scope of this report. Moreover, other EGOS characteristics should be more influenced by the neurons. For instance, it is known that the low frequency noise of the transistor is correlated to the number of viable cells anchored to the device. In silicon FETs[33] and OFETs,[34] the noise power spectral density is colored in the presence of viable cells, while it become 1/f when cells are either dead or absent. Also changes in power spectral density of OECTs in contact with living central nervous system tissue have been reported.[21] Whether also the STP response of EGOS will be affected by the viability (dead or alive) of cells would be possibly a subject of further investigations.

## 4. METHODS

*EGOS on Si/SiO$_2$ substrates*. The EGOSs were fabricated according to a process already described.[2, 3, 5] In brief, for measurements in air we used a bottom-gate electrode configuration with the gate electrode consisting of a highly-doped (resistivity ~10$^{-3}$ Ω.cm) n-type silicon wafer covered with a thermally-grown, 200 nm thick, silicon dioxide. We fabricated specific test patterns (channel length L=1 to 50 μm and width W=1000 μm) for electrical measurements in liquid with contact pads far from the electrodes (Fig. S1, see supporting information). Au NPs (10 ± 1 nm in diameter) were deposited on the substrate with a density of about 1-5x10$^{10}$ NP/cm$^2$ (supporting information, Fig. S1) followed by the thermal evaporation of a 40 nm thick pentacene (see SI). For the measurements in liquid, a Teflon pool was placed on the sample (supporting information Fig. S1) to confine liquid on top of the devices, and a Pt wire was directly immersed in the electrolytic solution and used as the gate electrode. The current-voltage characteristics were measured with an Agilent 4155C semiconductor parameter analyzer. For STP measurements, the input pulses were delivered by a pulse generator (Tabor 5061) or a Keysight B2902 source measure unit. The electrodes were contacted with a shielded micro-manipulator probe station in the dark. The electrolyte was a saline aqueous solution, i.e. deionized water with 0.1 M of NaCl. This ionic strength was chosen to be similar to physiological conditions.

*EGOS on quartz substrate*. For measurements with neuron cells, we fabricated EGOS on quartz (Fig. S2, supporting information), a transparent substrate being required for optical inspection of the deposition and adhesion of the stem cells and their differentiation into neurons. The layout geometry features channel length (L) of 15 $\mu$m and a channel width (W) close to 40000 $\mu$m. As a result, the geometrical ratio (W/L) is around 2600. Fabrication of EGOS onto these substrates followed the above-described protocol (see SI).

*STP measurements*. For the STP measurements, we used the EGOS as a two-terminal device. The source (S) and gate (G) electrodes were connected together and used as the input terminal of the device, while the drain (D) was used as the output terminal. Sequences of spikes with amplitude $V_p < 0$, duration(1-100 ms) and different frequencies were applied at the G/S input and the output drain current measured.

*Cell medium confinement and protocol for neurons growth.* Aiming at the electrical characterization of these devices during the cell growth, a system capable of confining cell medium on top of the interdigitated Au contacts while allowing at the same time optical microscopy to inspect cell growth and behavior is required. To this end we fixed the device on a metal plate with an opening below the interdigitated electrode area to permit access by an inverted microscope. The cell culture medium was contained in a polypropylene pool of 1 mL volume fixed on top of the device with a silicon rubber gasket (see Fig. S4, SI).[20]

Two neural cell lines were used in this work: human neuroblastoma SH-SY5Y cells (ECACC 94030304; Sigma–Aldrich, St. Louis, MO, USA) and NE-4C cells (ATTC No.: CRL-2925) as model systems. Details of the cells adhesion, growth and differentiation are given in the supporting information, following protocols previously described Refs. [20], [35] and [36]. Samples were imaged with Nikon Eclipse 80i microscope equipped for fluorescence analysis.

**Supporting Information**. Details on the EGOS fabrication, NP synthesis, cell culture, growth and differentiation, data on hole mobilities and AFM images are available free of charge in the supporting information.

**Corresponding Author**

£ Corresponding author : dominique.vuillaume@iemn.univ-lille1.fr

**Author contribution.**

SD, ML and SC did the electrical measurements on EGOS in electrolyte and with neurons. SD, AK, MM and TC did the experiments on the organic synapstor in air. ST and MB did the cell growth, differentiation and optical measurements. SD and DG fabricated the devices. SC, TC, FB and DV supervised the project. All the authors discussed the results and contributed to write the paper.

**Present Addresses**

† SD is now at Biophyresearch, Fuveau (France). SC is now at Institut de Ciència de Materials de Barcelona (ICMAB-CSIC), Universitat Autònoma de Barcelona (Spain). ST is now at ISOF-CNR, Bologna (Italy), AK is now at CEA-LITEN and INAC, Grenoble (France)

**ACKNOWLEDGEMENTS**.

This work has been financially supported by the EU 7th framework program under grant agreement n° 280772, project "Implantable Organic Nanoelectronics" (iONE-FP7). We thank F. Alibart and S. Lenfant for helpful support.

**NOTES & REFERENCES**

Russo, Tough and adhesive nanostructured calcium phosphate thin films deposited by the pulsed plasma deposition method, RSC Adv., 5 (2015) 78561-78571.

**Graphical abstract.**

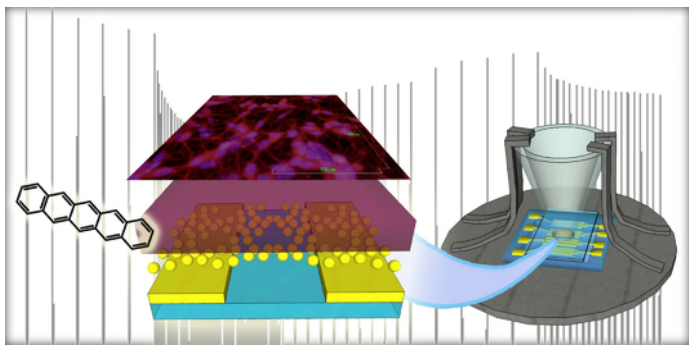

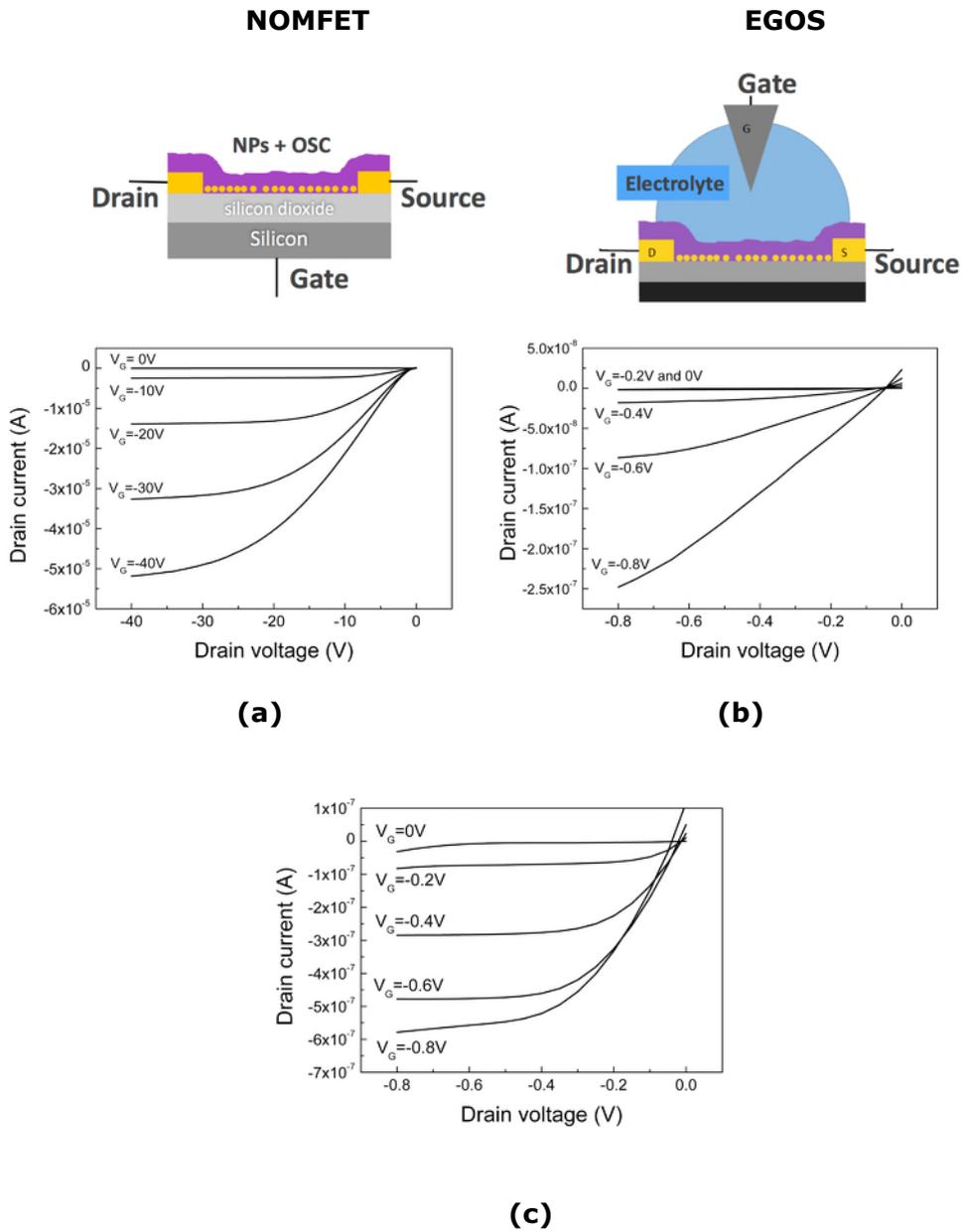

**Figure 1**. (Top, left) Schematics of a NOMFET and (top, right) EGOS. Output curves $I_D$-$V_D$ of (a) NOMFET (L= 5 μm) on Si/SiO$_2$ substrate in air, (b) EGOS (L = 5 μm) on Si/SiO$_2$ substrate and (c) EGOS (L = 15 μm) on quartz. Both EGOSs

on Si/SiO$_2$ and quartz substrates are measured in 0.1 M of NaCl in deionized water.

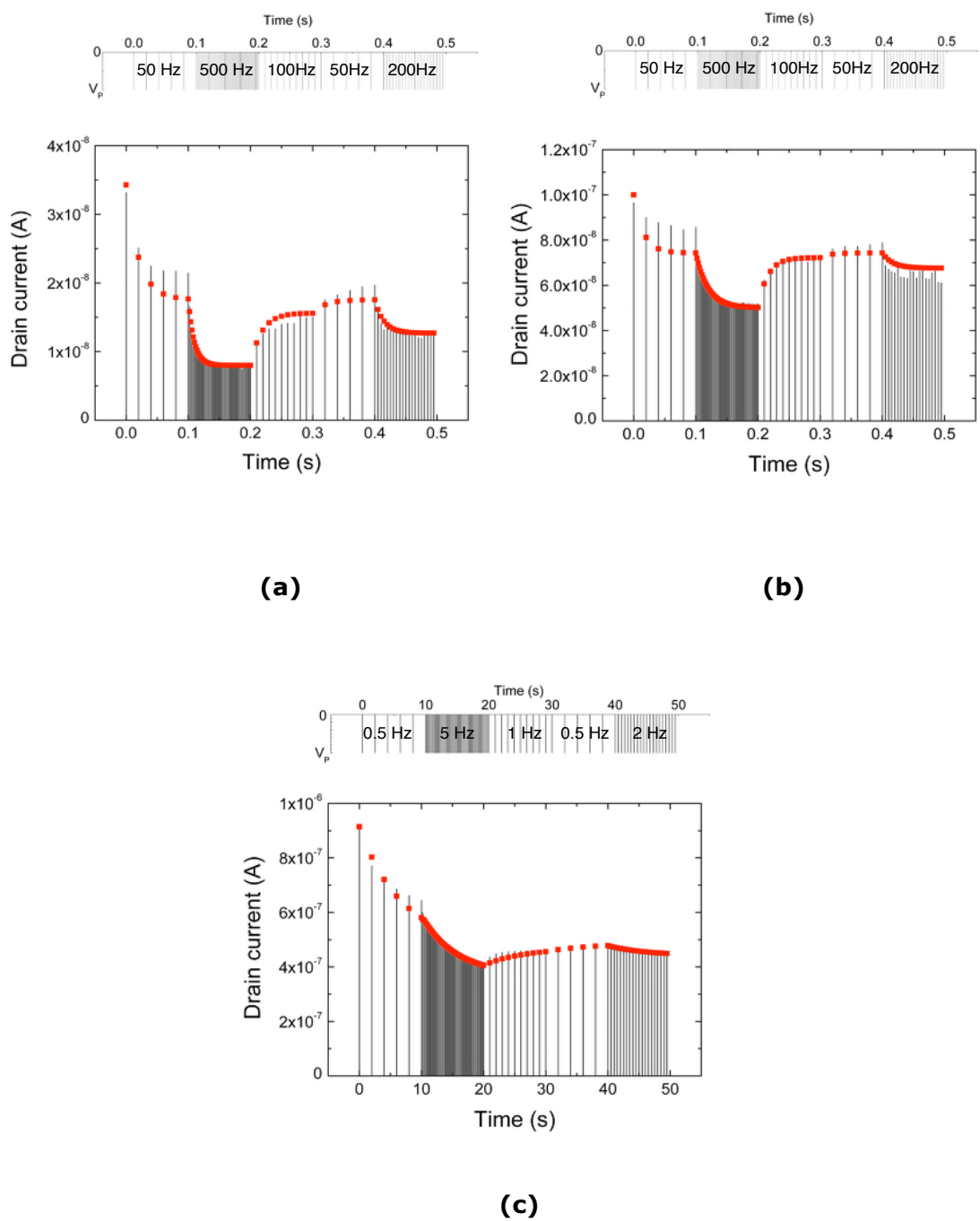

**Figure 2**. (a) STP response of EGOS (L=1 μm, on Si/SiO$_2$ substrate) in saline aqueous solution (V$_p$= -50 mV), the STP time constant is 10 ms according to the fit (red dots) with the analytical model of Ref. [2], the STP amplitudes $\Delta I/I$ is 0.32; (b) same device in air (Si bottom gate) with $V_p$ = -1V, best-fit STP time constant is 18 ms, the STP amplitudes $\Delta I/I$ is 0.28. In both cases, series of spikes at 50Hz (x5), 500Hz (x50), 100Hz (x10), 50Hz (x5) and 200Hz (x20), and 1 ms spike width, are applied (as shown above the figures). (c) STP response of EGOS on quartz (L=15 μm) in saline aqueous solution (($V_p$= -500mV). Best-fit of STP time constant is 6.5s. Spikes are applied at 0.5Hz (x5), 5Hz (x50), 1Hz (x10), 0.5Hz (x5) and 2Hz (x20), spike width is 100 ms.

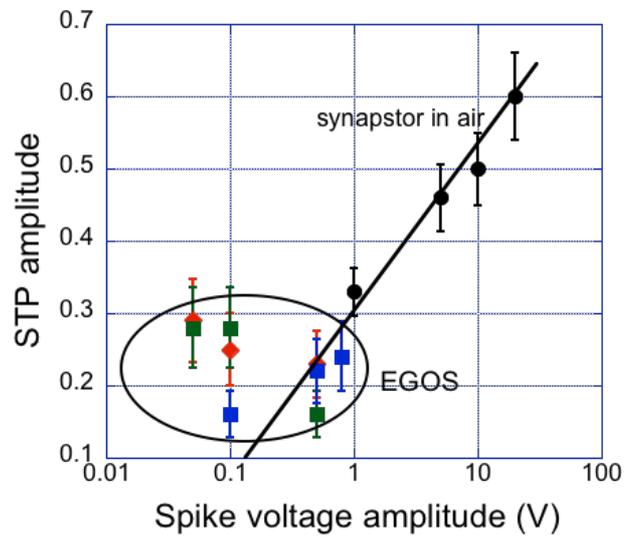

**Figure 3**. Evolution of the STP response amplitude for organic synapstor operated in air and EGOS versus absolute value of the spike amplitude. The line is a guide for eyes.

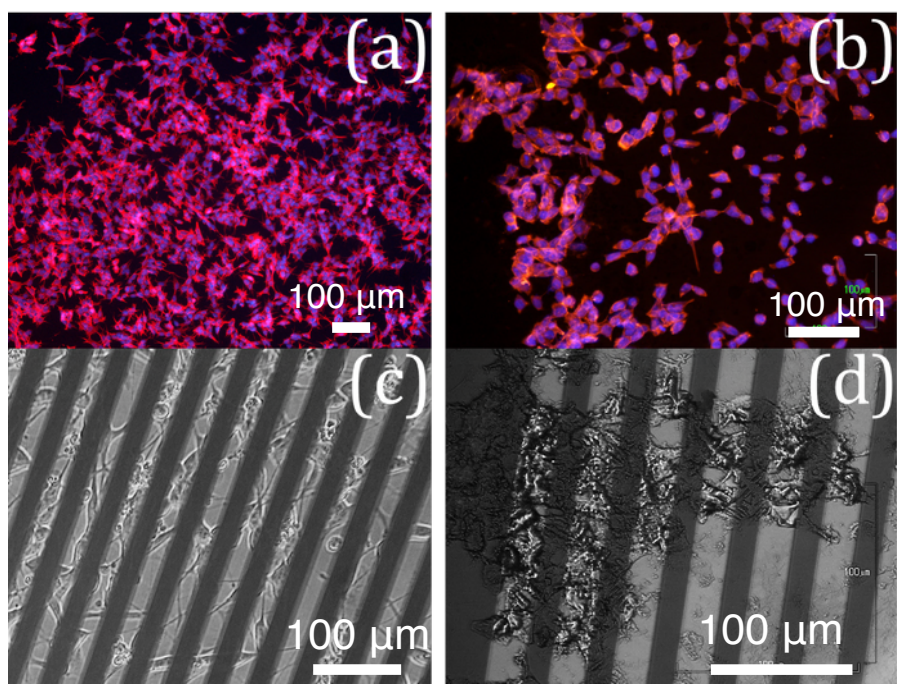

**Figure 4**. SH-SY5Y cells grown on quartz (a) and test-pattern (interdigitated Au electrodes) (c) coated by pentacene/NP thin-film. NE-4C cells grown on quartz (b) and test-pattern (d) coated by pentacene/NP thin-film. Immuno-fluorescence images of SH-SY5Y (a) and NE-4C (b) feature DAPI (blue color, labeling nuclei of the cells) and Alexa Fluor 594 (red) labels associated to the staining of β-III tubulin. (c) and (d) are the respective optical bright field images.

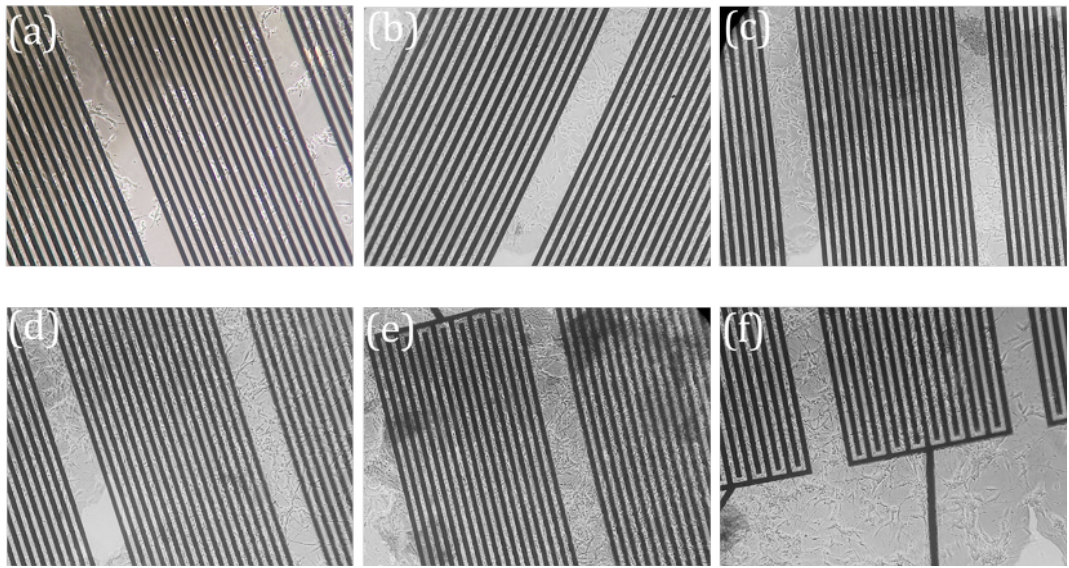

**Figure 5**. Optical bright field images (980 μm x 740 μm) of SH-SY5Y cells on EGOS at day 0 (a), 3 (b), 4 (c), 5 (d), 6 (e) and 7 (f). Day 0 stands for cell seeding, whereas day 3 means BDNF addition replacing retinoic acid, which is photosensitive.

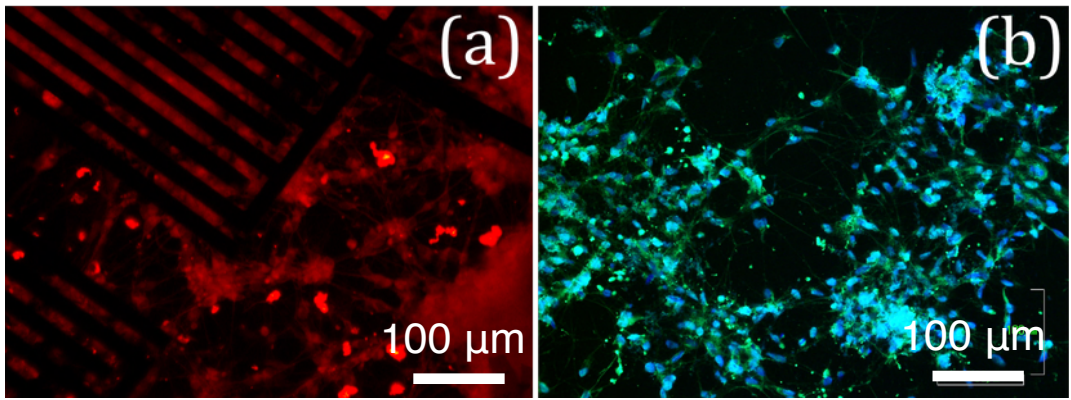

**Figure 6**. Immuno-fluorescences images of SH-SY5Y grown and differentiated on EGOS. (a) Red color is yielded by Alexa Fluor 594 immuno-label staining β-III tubulin, whereas (b) blue and green colors stand for DAPI dye (cell nuclei) and FITC immuno-label staining MAP2 (microtubule associated protein).

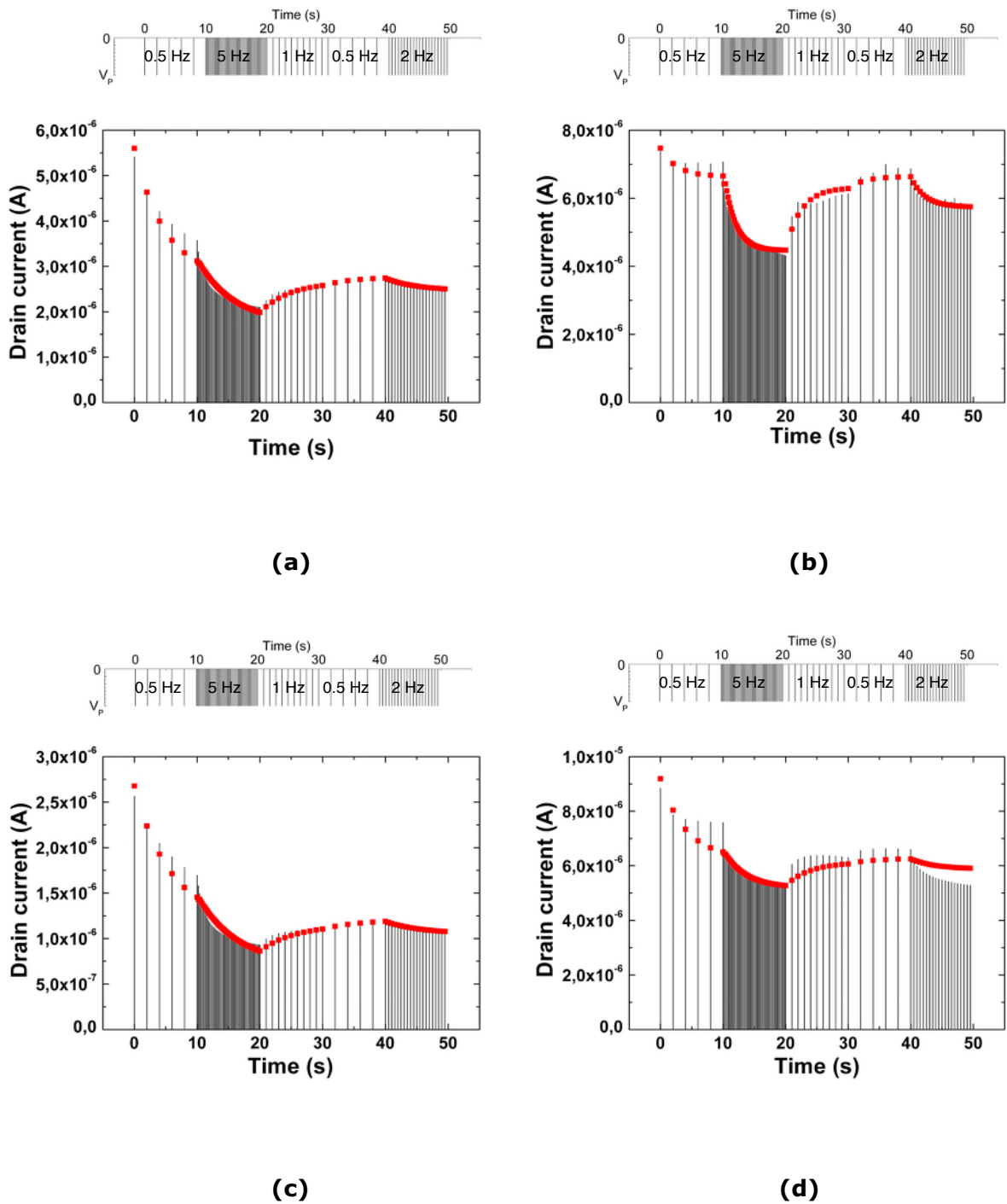

**Figure 7**. STP recorded on two EGOS devices with a series of spikes at 0.5Hz (x5), 5Hz (x50), 1Hz (x10), 0.5Hz (x5) and 2Hz (x20), spike width 100 ms,

Vp=-0.8V (shown above the figures). (a) STP at day 0 and (b) and day 1 (cell adhesion) for the device A, the STP time constants (given by the fits, red dots, with the analytical model of Ref.[ 2]) and STP amplitudes ΔI/I are 4.7s/0.43 and 2.6s/0.35, respectively. (c) STP at day 0 and (d) day 6 (cell differentiation) for device B, the STP time constants and STP amplitudes ΔI/I are 5.5s/0.44 and 4s/0.3, respectively. Note that 0.8V is the maximum voltage that we can applied (see Fig. 1-b) avoiding water electrolysis and faradaic current. Albeit a degradation of the devices (decrease of the current) has been observed for DC measurements over a long period (days), we have not observed any significant degradation of the EGOS behavior during the duration of the spike measurements with current remaining in the µA range. Note also that cells did not suffer because of the potential chosen.

# Electrolyte-gated organic synapse transistor interfaced with neurons

Simon Desbief*, Michele di Lauro#, Stefano Casalini#, David Guerin*, Silvia Tortorella+, Marianna Barbalinardo+, Adrica Kyndiah+, Mauro Murgia+, Tobias Cramer+§, Fabio Biscarini# and Dominique Vuillaume*.

*Institute for Electronics Microelectronics and Nanotechnology (IEMN), CNRS, Avenue Poincaré, F-59652cedex, Villeneuve d'Ascq, France.

# Life Science Dept., Università di Modena e Reggio Emilia, Via Campi 103, 41125 Modena, Italy.

+ Instituto per lo Studio dei Materiali Nanostrutturati (ISMN), CNR, Via P. Gobetti 101, 40129 Bologna, Italy.

§ University of Bologna, Dept. of Physics and Astronomy, Viale Berti Pichat 6/2, 40127 Bologna, Italy.

*Corresponding author : dominique.vuillaume@iemn.univ-lille1.fr*

# SUPPORTING INFORMATION

**MATERIALS AND METHODS**

***EGOS fabrication.***

*EGOS on Si/SiO$_2$ substrates*. The highly-doped (resistivity ~$10^{-3}$ Ω.cm) n-type silicon wafer was covered with a thermally-grown, 200 nm thick, silicon dioxide (deposition T=1100°C during 135 min in a dry oxygen flow, followed by a post-oxidation annealing at 900°C during 30 min under a nitrogen flow to reduce the presence of defects into the oxide). Metal electrodes (titanium/gold (20/80 nm)) were deposited on the substrate by vacuum evaporation and lift-off, patterned by e-beam lithography for linear source and drain gold electrodes (channel length L=1 to 50 µm and width W=1000 µm). To anchor gold NPs on the surface, the SiO$_2$ gate dielectric was cleaned by UV ozone treatment for 30 min and immediately functionalized with a self-assembled monolayer (SAM)[1] by immersion in a 1µL/mL solution of (3-aminopropyl)-trimethoxysilane (APTMS) in methanol for 24h.[2] The samples were cleaned thoroughly with methanol and then deionized water and immediately dipped in a water-dispersion of citrate-stabilized Au-NPs for 24h. We synthesized the colloidal solutions of citrate-capped Au NPs (10 ± 1 nm in diameter) by the Turkevich method (reduction of chloroauric acid by citrate in water).[3] This procedure yields an array of not coalesced NPs with a density of about 1-5x10$^{10}$ NP/cm$^2$ (Fig. S1). Then, silanization reaction with octadecyltrichlorosilane (OTS) was carried out as reported elsewhere to improve the performance of the devices in terms of hole mobility.[4] The silanization reaction was carried out in a glovebox under nitrogen atmosphere but non-anhydrous solvents were used to favor hydrolysis of -SiCl$_3$ functions. The freshly substrate was immersed for 2h in a 10$^{-3}$ M solution of OTS in a mixture of n-hexane and dichloromethane (70:30 v/v). The device was rinsed thoroughly by sonication in dichloromethane (2 times) then blown with dry nitrogen. Finally, we thermally evaporated a 40 nm thick pentacene film at a rate of 0.1 Å/s. with the substrate kept at 60°C. For the measurements in liquid, a Teflon pool was placed on the sample (Fig. S1) to confine liquid on top of the devices, and a Pt wire was directly immersed in the electrolytic solution and used as the gate electrode.

Figure S1-a shows a SEM image (1.7 µm x 1.4 µm) of the NPs deposited between the source and drain electrodes before pentacene thin film deposition, the density is ca. 4.5x10$^{10}$ NPs/cm$^2$. Fig. S1-b show the Teflon

pool used to measure the electrical characteristics of the EGOS and the corresponding device layout.

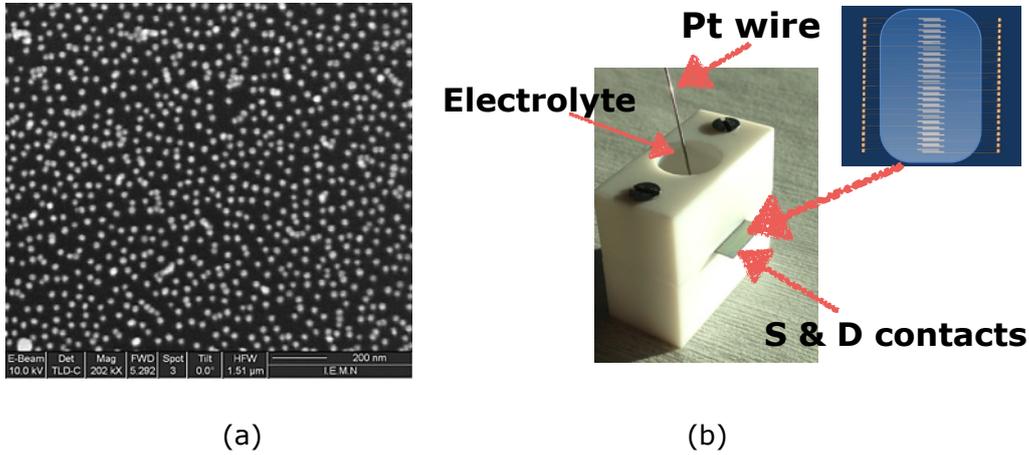

(a)                  (b)

***Figure S1****. (a) SEM image (1.7 µm x 1.4 µm) of the NPs deposited between the source and drain electrodes before pentacene deposition. The density is ca. $4.5 \times 10^{10}$ NPs/cm$^2$. (b) Picture of the EGOS measurement system (Teflon pool) and layout of the devices.*

*EGOS on quartz substrate*. We used quartz substrates with a thickness of 800(±50)µm and a roughness of ca. 2 nm. The interdigitated Au electrodes are 40nm thick with few nm of Ti layer in order to guarantee a robust adhesion. The layout geometry features channel length (L) of 15 µm and a channel width (W) close to 40000 µm. As a result, the geometrical ratio (W/L) is around 2600. Fabrication of EGOS onto these substrates followed the above-described protocol, with the exclusion of the channel silanization with OTS. Fig. S2 shows a picture of the fabricated EGOS on quartz substrate. Figure S3 shows a SEM image of the NPs network on quartz (before pentacene deposition).

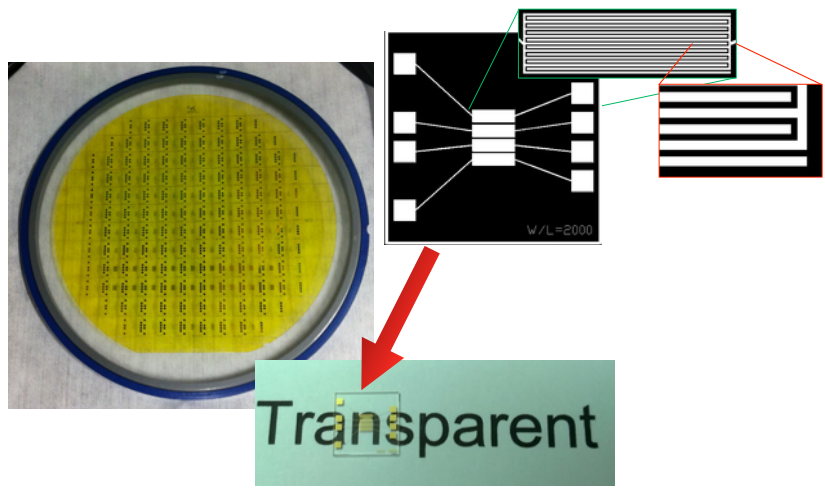

*Figure S2*. Images of the processed quartz substrates, and layout of EGOS on quartz

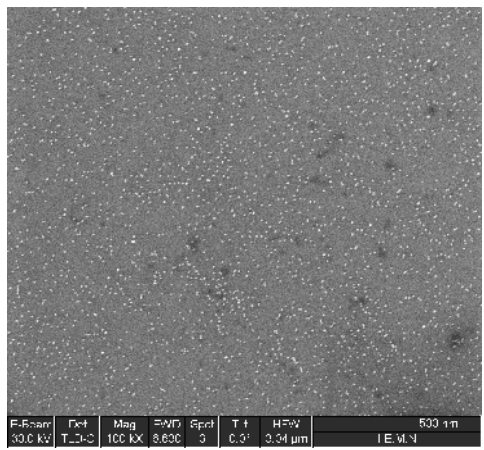

*Figure S3.* SEM image of the NP network (on a piece of Si/SiO$_2$ processed simultaneously with the quartz substrate, since SEM image is not possible on insulating quartz substrate. We assume that the same density is achieved on both surfaces having received strictly the same surface treatments). SEM images 3 µm x 3 µm. The NP density is ca. 4.4x10$^{10}$ NPs/cm$^2$

**CELL CULTURE ON EGOS**

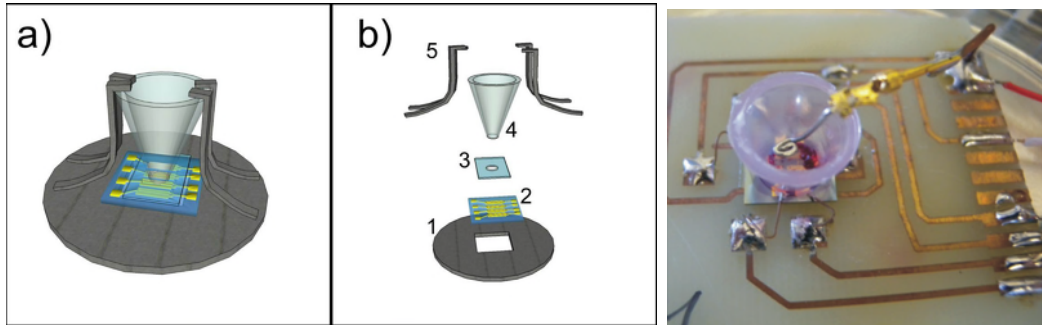

*Figure S4. Home-made system (Ref. 5) completely assembled (a) and disassembled (b) featuring the single components: 1. Metal basis, 2. Test pattern, 3. Rounded and flat pool, 4. Truncated conical poly-propylene pool and 5. Metal arms. (c) Picture of the system.*

Cells were growth at 37°C in humidified atmosphere (95% humidity, 5% $CO_2$) at density of $2.0 \times 10^5$ cells per $cm^2$ in complete medium (CM) composed by Minimum Essential Medium (MEM) containing 10% of fetal bovine serum, 4 mM of L-glutamine, 40 µg/mL of gentamicin and 2.5 µg/mL of amphotericin). Cell differentiation was induced by replacing CM with the differentiation medium (DM) composed of Dulbecco's Modified Eagle's Medium DMEM-F-12 Ham (1:1), Insuline Transferrin Selenium (ITS) 1X, 4 mM glutamine, 40 mg/ mL gentamicin and 2.5 mg/mL amphotericin, containing 10 µM of all-trans retinoic acid (RA) as previously described Refs. 5, 6 and 7.

The samples were fixed for 15 min in 4% paraformaldehyde in phosphate buffered saline (PBS; pH 7.4) and permeabilized with 0.1% Triton X100 (Sigma) in PBS for 5min. They were incubated in 1% bovine serum albumin in PBS to block non-specific binding sites. Cell adhesion was characterized by Actin Cytoskeleton and Focal Adhesion Staining kit (FAK100, Millipore, USA). Neuronal cells differentiated by RA for 7 days were fixed and processed with the primary antibody anti-Tubulin β-III (Sigma T2200) and anti-MAP2 microtubule-associated protein (Sigma M3696) as neuronal marker, with the aim to check the neuronal differentiation on substrates.

**HOLE MOBILITY**

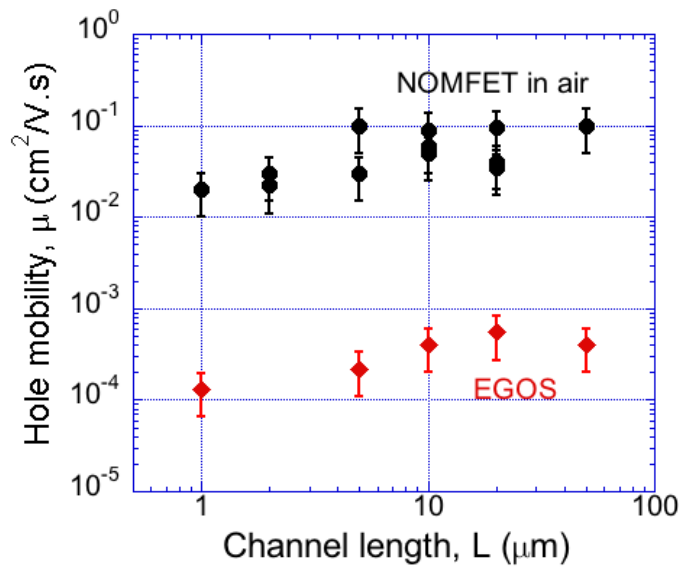

*Figure S5.* Hole mobility in saturation versus channel length for devices measured in air (NOMFET configuration, Si bottom gate) and in EGOS configuration (electrolyte gated).

**AFM IMAGES**

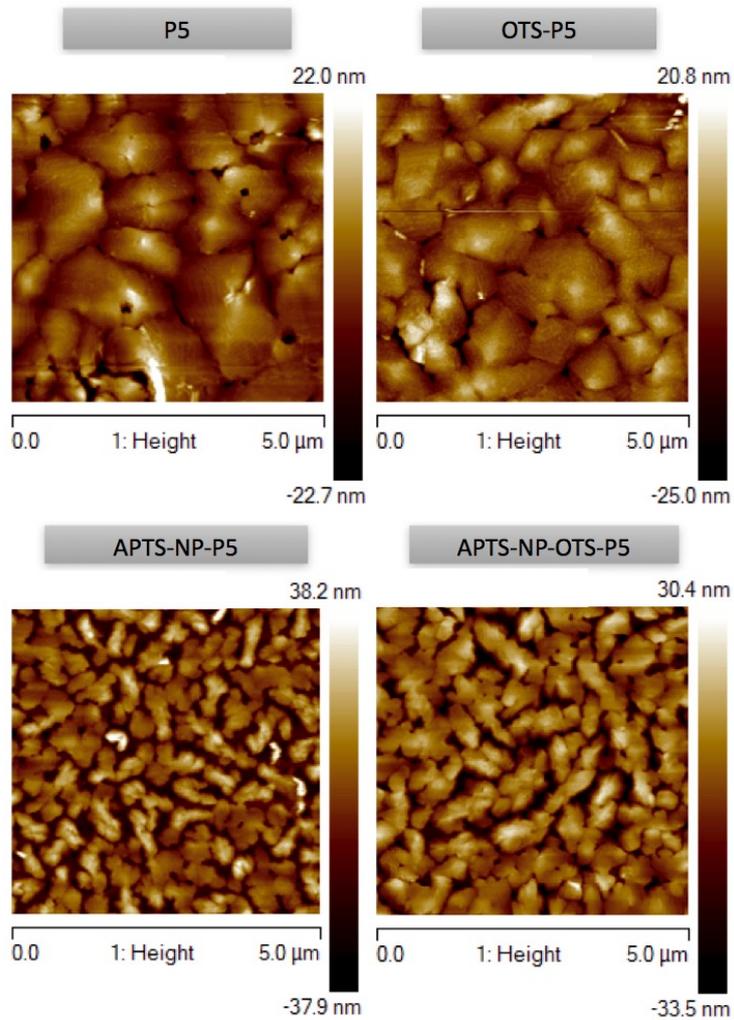

***Figure S6.*** *Tapping mode-AFM images of the pentacene films for 2 OFETs (top) without NPs: pentacene on SiO$_2$ (left) and pentacene on OTS-treated SiO$_2$ (right); and 2 organic synapstor (bottom) with NPs: APTMS-NP-P5 (left) and APTMS-NP-OTS-P5 (right) structures (Ref. 4). Image analysis (PSD : power spectral density) gives the following average grain size : 1.25 µm, 1.67 µm, 0.55 µm and 0.83 µm, respectively. The rms roughnesses are : 15 nm, 8 nm, 27 nm and 15 nm, respectively.*